\documentclass[manuscript]{aastex61}

\usepackage{multirow}
\usepackage{booktabs}
\usepackage{rotating}

\hypersetup{linkcolor=red,citecolor=blue,filecolor=cyan,urlcolor=magenta}

\newcommand{\wise}{{\textit{WISE~}}}
\newcommand{\herschel}{{\textit{Herschel~}}}
\newcommand{\um}{$\mu m$~}

\shorttitle{SCUBA-2 follow-up of dust-obscured quasars}
\shortauthors{Fan et al.}

\begin{document}

\title{The SCUBA-2 850 \um\ follow-up of WISE-selected, luminous dust-obscured quasars}

\correspondingauthor{Lulu Fan}
\email{llfan@sdu.edu.cn}

\author{Lulu Fan}
\affil{Shandong Provincial Key Lab of Optical Astronomy and Solar-Terrestrial Environment, Institute of Space Science, Shandong University, Weihai, 264209, China}

\author{Suzy F. Jones}
\affiliation{Department of Space, Earth and Environment, Chalmers University of Technology, Onsala Space Observatory, SE-439 92 Onsala, Sweden}

\author{Yunkun Han}
\affiliation{Yunnan Observatories, Chinese Academy of Sciences, Kunming, 650011, China}

\author{Kirsten K. Knudsen}
\affiliation{Department of Space, Earth and Environment, Chalmers University of Technology, Onsala Space Observatory, SE-439 92 Onsala, Sweden}



\begin{abstract}

Hot dust-obscured galaxies (Hot DOGs) are a new population recently discovered in the \wise All-Sky survey. Multiwavelength follow-up observations suggest that they are luminous, dust-obscured quasars at high redshift. Here we present the JCMT SCUBA-2 850 $\mu m$ follow-up observations of 10 Hot DOGs. Four out of ten Hot DOGs have been detected at $>3\sigma$ level. Based on the IR SED decomposition approach, we derive the IR luminosities of AGN torus and cold dust components. Hot DOGs in our sample are extremely  luminous with most of them having $L_{\rm IR}^{\rm tot}>10^{14} L_\odot$. The torus emissions dominate the total IR energy output.  However, the cold dust contribution is still non-negligible, with the fraction of the cold dust contribution to the total IR luminosity $(\sim 8-24\%)$  being dependent on the choice of torus model. The derived cold dust temperatures in Hot DOGs are comparable to those in UV bright quasars with similar IR luminosity, but much higher than those in SMGs. Higher dust temperatures in Hot DOGs may be due to the more intense radiation field caused by intense starburst and obscured AGN activities. Fourteen and five submillimeter serendipitous sources in the 10 SCUBA-2 fields around Hot DOGs have been detected at $>3\sigma$ and $>3.5\sigma$ levels, respectively. By estimating their cumulative number counts, we confirm the previous argument that Hot DOGs lie in dense environments. Our results support the scenario in which Hot DOGs are luminous, dust-obscured quasars lying in dense environments, and being in the transition phase between extreme starburst and UV-bright quasars.

\end{abstract}





\section{Introduction}

Massive galaxies have been thought to co-evolve with their central supermassive black holes \citep{alexander2012,kormendy2013}.  In the co-evolution scenario, starburst galaxies, such as ultraluminous infrared galaxies (ULIRGs; Sanders \& Mirabel 1996) and sub-millimeter galaxies (SMGs; Blain et al. 2002; Casey et al. 2014), ultraviolet (UV) bright quasars (QSOs) and massive early-type galaxies (ETGs) have been proposed to form an evolutionary sequence \citep{sanders1988,hopkins2008}. Luminous dust-obscured QSOs represent a brief transition phase from starburst galaxies to unobscured QSOs. It is important to understand this obscured phase, especially at $z\sim2-3$ when both cosmic star formation and black-hole growth reach their peaks \citep{reddy2008,brandt2015}. However, it is not easy to select large samples of obscured QSOs in the normal UV/optical and X-ray surveys due to dust obscuration. Dust  absorbs the UV and optical radiations from star formation and the black-hole accretion disk, and re-emits at far-IR/(sub-)mm wavelengths, making them be very luminous in the IR but faint at UV and optical wavelengths. 

NASA's {\it Wide-field Infrared Survey Explorer} (\wise; Wright et al. 2010) performed an all-sky  survey with images in four mid-IR bands ({\textit W1},{\textit W2},{\textit W3} and {\textit W4}) at wavelengths of 3.4 $\mu m$, 4.6 $\mu m$, 12 $\mu m$ and 22 $\mu m$, providing the sensitivity to detect luminous dust-obscured QSOs. One of the primary science objectives for \wise is to identify the most luminous ULIRGs in the Universe. In order to achieve this goal,  a new color-selected method (so-called {\textit W1}{\textit W2}-dropouts) with all \wise four bands has been recently introduced and proven to be very successful \citep{eisenhardt2012,wu2012}. This method searched the luminous dust-obscured galaxies, which are prominent in the \wise 12 or 22 $\mu m$ bands, and  faint or undetected in the 3.4 and 4.6 $\mu m$ bands.  Spectroscopic follow-up of about 150  galaxies selected with this method have shown that most of them are at redshift from 1 to 4 \citep{tsai2015,assef2015}. Given bright {\textit W3/W4} bands and high redshift, these galaxies should have very high mid-IR luminosities. 

Various follow-up studies have provided additional aspects on understanding the physical properties of this population.  Together with \wise photometry, continuum measurements at longer wavelengths with the {\it Herschel} Space Observatory \citep{pilbratt2010}  PACS (Photodetector Array Camera and Spectrometer; Poglitsch et al. 2010) observations at 70 and 160 $\mu m$ and SPIRE (Spectral and Photometric Imaging REceiver; Griffin et al. 2010) observations at 250, 350, and 500 $\mu m$ (PI: P.R.M. Eisenhardt),   the CSO (Caltech Submillimeter Observatory ) SHARC-II and  Bolocam observations at 350, 450, 850 and 1100 $\mu m$ \citep{wu2012} and the SCUBA-2 (Submillimeter Common-User Bolometer Array) 850 $\mu m$ observations \citep{jones2014} provide the full IR spectral energy distributions (SEDs) of tens of these galaxies, which suggest that (1)  they are indeed hyper-luminous with $L_{\rm bol}>10^{13}L_\odot$ and even extremely luminous with $L_{\rm bol}>10^{14}L_\odot$ \citep{tsai2015}; (2) their IR luminosities are dominated by hot dust radiation which is heated by active galactic nuclei (AGN) instead of star formation. The mid-IR bright criterion is very similar to that of dust-obscured galaxies (DOGs; Dey et al. 2008). Following the terminology of \citet{wu2012}, these galaxies have been referred as hot, dust-obscured galaxies (Hot DOGs). The detailed IR SED decomposition confirmed that Hot DOGs have both extreme AGN and starburst activities and hot dust emission from AGN torus dominates the IR energy output \citep{fan2016b}.  Recent X-ray observations of several Hot DOGs with XMM-Newton, Chandra and NuSTAR \citep{stern2014,assef2016,piconcelli2015,ricci2017} revealed that they are likely heavily dust-obscured, possibly Compton-thick AGN.

Both extreme AGN and starburst activities in Hot DOGs are likely triggered by major mergers,  which has been suggested by the recent studies of the high resolution HST WFC3 NIR imaging (Fan et al. 2016a; but also see Farrah et al. 2017 for an opposite conclusion) and  ALMA CO line imaging (Fan et al. 2017 in prep).  \citet{wu2017} measured the supermassive black hole (SMBH) masses via broad H$_\alpha$ emission lines, using Keck/MOSFIRE and Gemini/FLAMINGOS-2. They found substantial corresponding SMBH masses for these Hot DOGs ($\sim10^9 M_\odot$), and their derived Eddington ratios are close to unity (see also the similar results in Lonsdale et al. 2015 for a radio-selected Hot DOG sample).  A recent ALMA [CII] observation of the most luminous Hot DOG W2246$-$0526 revealed an extended and highly turbulent ISM,  arguing that  it is blowing out its ISM isotropically \citep{diaz-santos2016}. Given the major-merger triggering scenario, high luminosities, high SMBH masses, high Eddington ratios and strong feedback, Hot DOGs, as luminous dust-obscured QSOs, are suggested to be in a key transition phase of massive galaxy evolution, linking starbursts and luminous unobscured quasars \citep{wu2012,bridge2013,diaz-santos2016,fan2016b,wu2017}. 

Despite the dominance of AGN torus emission, cold dust emission related to star formation is non-negligible \citep{wu2012,fan2016b}. Constraining the contribution of cold dust emission will be very important to help understanding the co-growth of galaxy and SMBH during the dust-obscured phase. In this work, we present James Clerk Maxwell Telescope (JCMT) SCUBA-2 850 $\mu m$ observations of ten Hot DOGs. With these long wavelength measurements, we try to constrain the contribution of cold dust emission compared to the AGN torus emission and study the SMG environment around Hot DOGs. In Section \ref{sec:data}, we describe the sample selection, observations and data analysis. We introduce the IR SED decomposition method with two different model sets in Section \ref{sec:sedfitting}. In Section \ref{sec:res}, we present the results of IR SED fitting and the environmental studies around Hot DOGs. We discuss and summarize the results in Section \ref{sec:dis} and Section \ref{sec:sum}, respectively. Throughout this work we assume a standard, flat ${\rm \Lambda}$CDM cosmology \citep{komatsu2011}, with $H_0 = 70$ km~s$^{-1}$, $\Omega_M = 0.3$, and $\Omega_\Lambda = 0.7$.

\section{Data and sample selection}\label{sec:data}

\subsection{Sample selection}\label{subsec:sample}

The ten Hot DOGs reported here are selected from the \wise All-Sky Source catalog, which provides  point-spread function-fitting
magnitudes and uncertainties of four mid-IR bands ({\textit W1},{\textit W2},{\textit W3} and {\textit W4})
at wavelengths of 3.4 \um , 4.6 \um, 12 \um and 22 \um \citep{wright2010,cutri2013}. The basic idea of Hot DOGs
selection is to target those mid-IR luminous galaxies with faint or undetected {\textit W1}/{\textit W2}  bands,
and prominent  {\textit W3}/{\textit W4}  bands. The selected galaxies have  {\textit W1} $> 17.4$~ ($<34 \mu$Jy) and
either {\textit W3} $< 7.7$ ($>6.9$ mJy) and {\textit W2} $-$ {\textit W4} $ > 8.2$, or {\textit W3} $ < 10.6$ ($>1.7$ mJy)
and {\textit W2} $-$ {\textit W3} $> 5.3$ \citep{eisenhardt2012,wu2012}.

For our purposes of SCUBA-2 follow-up, we required that the selected Hot DOGs:
(1) had known spectroscopic redshift in the literature \citep{wu2012,tsai2015}.
(2) had \herschel PACS and SPIRE observations.
By using these criteria, we selected  nineteen Hot DOGs. Among them, ten Hot DOGs were randomly chosen to  observe by SCUBA-2
according to the JCMT queue-based flexible observing system.


\begin{sidewaystable}[h]
\centering
\caption{SCUBA-2 observations and photometry of the 10 Hot DOGs.}\label{tab:obs}
\begin{tabular}{lccccllc}
\hline
\hline
Source          & R.A.             & Decl.              & Redshift &  Exposure time   &  UT Date   &   $\tau_{225GHz}$   &  850$\mu m$   \\
Name            & (J2000)          & (J2000)            &          &  (hr)            &            &                     &    (mJy)      \\
\hline
  W0126$-$0529  &     01:26:11.96  &     $-$05:29:09.6  & 2.937  &  2.0 &   2015 Nov 18, Dec 25-26              &  0.09,0.10       & 15.9 $\pm$ 2.6 \\
  W0134$-$2922  &     01:34:35.71  &     $-$29:22:45.4  & 3.047  &  2.5 &   2015 Dec 25,27                      &  0.09,0.10,0.12  & $<5.3^{a}$     \\
                &                  &                    &        &      &   2016 Jan 7,Jun 4                    &                  &                \\
  W0248+2705    &     02:48:58.81  &     +27:05:29.8    & 2.210  &  1.0 &   2015 Dec 25                         &  0.09            & $<5.5^{a}$     \\
  W0422$-$1028  &     04:22:48.82  &     $-$10:28:32.0  & 2.227  &  2.0 &   2015 Nov 10,18                      &  0.10,0.09       & $<4.6^{a}$     \\
  W0757+5113    &     07:57:25.07  &     +51:13:19.7    & 2.277  &  2.0 &   2015 Oct 26-27                      &  0.10,0.09       & $<4.7^{a}$     \\
  W0859+4823    &     08:59:29.94  &     +48:23:02.3    & 3.245  &  1.5 &   2015 Oct 27-28                      &  0.09,0.08       & 7.0 $\pm$ 2.0  \\
  W1248$-$2154  &     12:48:15.21  &     $-$21:54:20.4  & 3.318  &  2.0 &   2015 May 12, Dec 25                 &  0.09            & $<4.7^{a}$     \\
  W1838+3429    &     18:38:09.16  &     +34:29:25.9    & 3.205  &  2.0 &   2015 Jun 1-2, Jul 16                &  0.11,0.12,0.09  & $<5.2^{a}$     \\
  W2201+0226    &     22:01:23.39  &     +02:26:21.8    & 2.877  &  2.0 &   2015 May 26                         &  0.10            & 17.2 $\pm$ 2.5 \\
  W2210$-$3507  &     22:10:11.87  &     $-$35:07:20.0  & 2.814  &  2.0 &   2015 Apr 25, May 31                 &  0.10, 0.09      & 12.3 $\pm$ 3.1 \\
                &                  &                    &        &      &   2015 Jun 17, Jul 16                 &                  &                \\
\hline
\end{tabular}
\parbox{150mm} {
\textbf{Notes.} \\
$^{a}$ 2$\sigma$ upper limits at 850 $\mu$m. \\
}
\end{sidewaystable}

\subsection{JCMT SCUBA-2 observations and photometry}\label{subsec:scuba2}

Observations of the ten Hot DOGs were obtained using the JCMT SCUBA-2 camera, scheduled flexibly in 2015A and 2015B semesters (Program ID: M15AI06 and M15BI016), from 2015 April to 2016 June. During the observations, the precipitable water vapor (PWV) was in the range 2.0-3.0 mm, corresponding to the zenith atmospheric optical depth at 225 GHz: $0.08<\tau_{225GHz}<0.12$ and the extinction coefficients at 450 \um and 850 \um:
$1.77<\tau_{450}<2.81$ and $0.35<\tau_{850}<0.53$ \citep{dempsey2013}. Even if both 450 \um and 850 \um  measurements were taken simultaneously, 450 \um data couldn't be used due to the high atmospheric opacity and noise levels.

All observations were carried out in SCAN mode using the so-called `DAISY' pattern which provides a 12-arcminute-diameter map with uniform exposure-time coverage in the central 3-arcminute-diameter region. The full width at half maximum (FWHM) of the main beam at 850 \um\ is about 14.5 arcsecond \citep{holland2013}. For seven out of ten targets, two hours of the total exposure time were spent, consisting of four separate 30-minutes-long scans. The remaining targets, W0134$-$2922, W0248+2705 and W0859+4823, had 150, 60 and 90 minutes exposure time, respectively. The calibration sources observed were Uranus and Mars, and also secondary calibrators from the James Clerk Maxwell Telescope (JCMT) calibrator list \citep{dempsey2013}. Calibrations were taken before and after the target observations.

The SCUBA-2 maps were reduced using  the Dynamic Iterative Map Maker (DIMM) within the STARLINK SubMillimeter User 
Reduction Facility (SMURF) data reduction package \citep{chapin2013}.  DIMM was used to produce a map for each 30-minutes-long scan
by using mapmaker command with the `Blank Field configuration' which is suitable for the faint point sources. Using the STARLINK PIpeline
for Combining and Analyzing Reduced Deduced Data (PICARD) package, all maps for each target were mosaicked, cropped to a 90-arcsecond-radius
circle, beam-match filtered with a 15-arcsecond FWHM Gaussian and calibrated with the flux conversion factors (FCFs). We adopted the FCFs value of
537\ Jy\ pW$^{-1}$\ beam$^{-1}$ for 850 \um data \citep{dempsey2013}.

In Table \ref{tab:obs}, we present the result of SCUBA-2 850 $\mu m$ photometry of the ten Hot DOGs . Four out of ten Hot DOGs (W0126$-$0529, W0859+4823, W2201+0226 and  W2210$-$3507) have been detected at $> 3\sigma$ significance and their flux densities and uncertainties are included in Table \ref{tab:obs}. The remaining six Hot DOGs (W0134$-$2922, W0248+2705, W0422$-$1028, W0757+5113, W1248$-$2154 and W1838+3429) are undetected. Their $2\sigma$ upper limits have been presented in Table \ref{tab:obs}. The photometric method presented here is identical to that in \citet{jones2014} towards a sub-sample of the same parent population of our targets. However, mainly due to the different observation conditions, our photometric results are slightly different from those in \citet{jones2014}. Our observations were in the range of JCMT Band 3 conditions while those of \citet{jones2014} were in the better conditions (Band 2). The average $1\sigma$ depths reach 2.6 mJy and 1.8 mJy for our observations and \citet{jones2014}, respectively.  Instead of 40\% detection rate in our observations, a higher detection rate (60\%) has been found in \citet{jones2014}.

\subsection{Ancillary data}\label{subsec:other}

In order to construct the full infrared SED of the ten SCUBA-2 observed Hot DOGs, we combined the public \wise and \herschel photometry. We retrieved the \wise {\textit W3} and {\textit W4} photometry from the ALLWISE Data Release \citep{cutri2013}. We used zero points of 29.04 and 8.284 Jy to convert {\textit W3} and {\textit W4} Vega magnitude to flux densities, respectively \citep{wright2010}.   All of the ten Hot DOGs in our sample had been observed by \herschel PACS \citep{poglitsch2010}  at 70 and 160 $\mu m$, and SPIRE \citep{griffin2010} at 250, 350 and 500 $\mu m$. Both PACS and SPIRE data were retrieved from the Herschel Science Archive (HSA)  and reduced using the Herschel Interactive Processing Environment (HIPE v14.2.1).  The flux densities and uncertainties of  PACS and SPIRE were measured using the identical method in \citet{fan2016b}. 

\section{IR SED fitting}\label{sec:sedfitting}

\begin{figure*}
\plotone{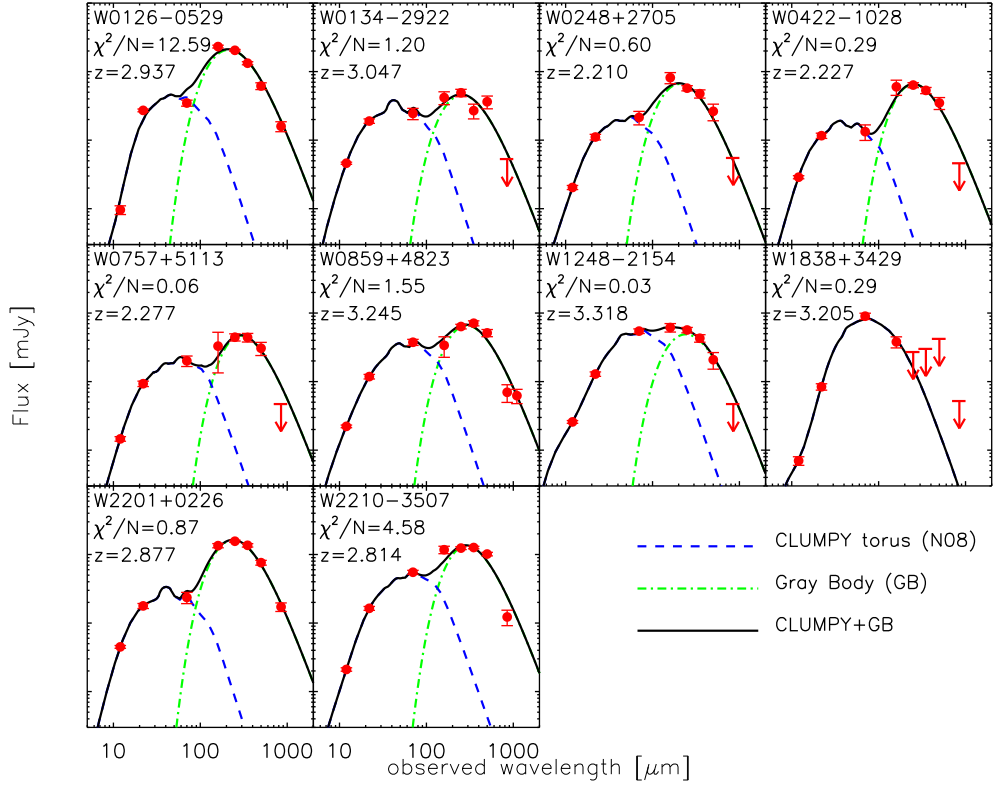}
\caption{Best-fit model SEDs with the CLUMPY+GB model for the ten SCUBA-2 observed Hot DOGs in our sample. The filled circles represent the observed data points. The downward arrows mark the upper limits of undetected bands. The solid line represents the combined CLUMPY+GB model. The dashed and dot-dashed lines represent the components of the clumpy torus and gray body in the CLUMPY+GB model, respectively.}\label{fig:fitn08}
\end{figure*}

\begin{figure*}
\plotone{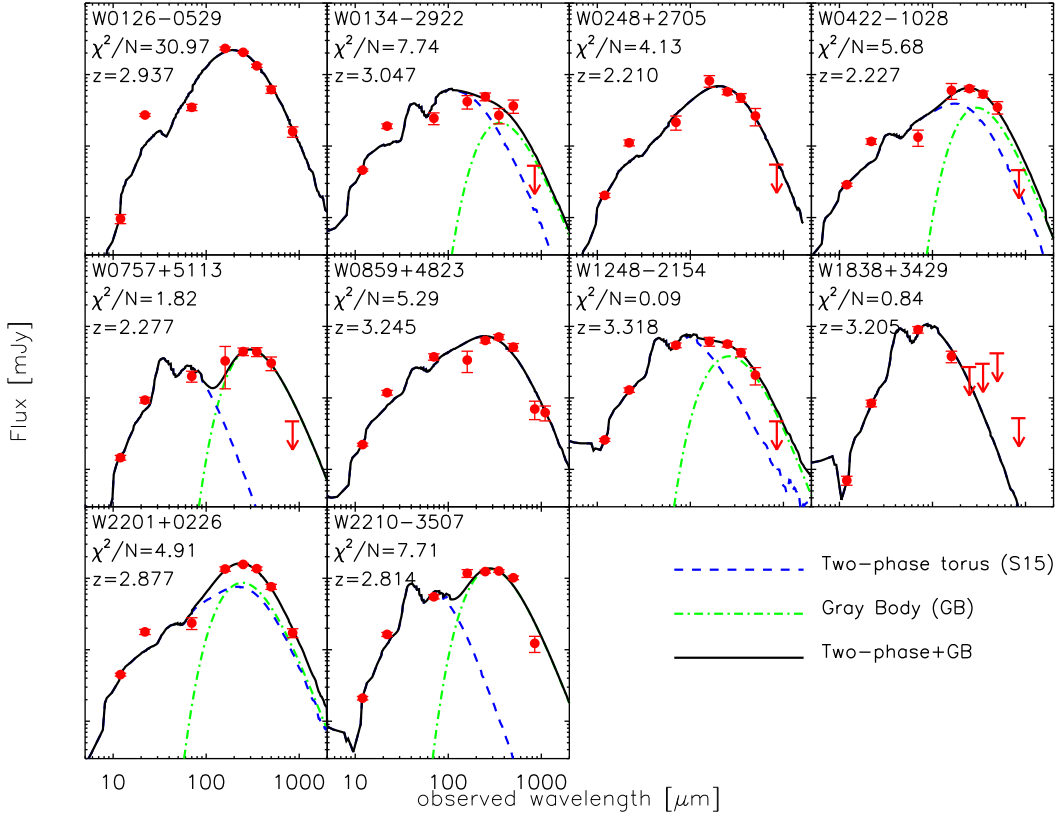}
\caption{Best-fit model SEDs with the Two-phase+GB model for the ten SCUBA-2 observed Hot DOGs in our sample. The filled circles represent the observed data points. The downward arrows mark the upper limits of undetected bands. The solid line represents the combined Two-phase+GB model. The dashed and dot-dashed lines represent the components of the two-phase torus  and gray body in the Two-phase+GB model, respectively. }\label{fig:fits15}
\end{figure*}

In this section, we describe how to fit the observed IR SEDs of the ten Hot DOGs. We assume that their IR SEDs are mainly distributed by two components. One is the hotter AGN heated dust emission and the other is colder young stellar population heated dust emission. It is obvious that the fitting results are dependent on which model is used for each component. For the AGN heated dust emission, we have employed two kinds of torus models: the CLUMPY torus model by \citet{nenkova2002,nenkova2008a,nenkova2008b}\footnote{www.clumpy.org} and the two-phase torus model by \citet{siebenmorgen2015}. For the young stellar population heated dust emission, we have employed a simple modified blackbody model (or a gray body, GB).

For the CLUMPY torus model, we use the newly calculated model database\footnote{http://www.pa.uky.edu/clumpy/models/clumpy\_models\_201410\_tvavg.hdf5/}. There are 6 free parameters in the CLUMPY torus model. We assume that the 6 free parameters have the uniform distributions within the following intervals: the number of clouds along a radial equatorial path $N_0= [1,15]$, the ratio of the outer to the inner radii of the toroidal distribution $Y=R_{\rm o}/R_{\rm d}=[5,100]$, the viewing angle measured from the torus polar axis $i =  [0,90]$, the index $q=[0,3]$ of the radial density profile  $r^{-q}$, the width parameter characterizing the angular distribution $\sigma= [15,70]$, and the effective optical depth of clumps $\tau_{\rm V} = [10,300]$.  Given a set of parameter values, a model SED can be calculated. There are in total 1,247,400 model SEDs in the database.

For the two-phase torus model \citep{siebenmorgen2015}, we use the provided SED library\footnote{http://www.eso.org/\~{}rsiebenm/AGN/}. Unlike the CLUPMY torus model, the two-phase torus model assumes that torus dust is not only distributed in a clumpy medium, but also a homogeneous disk.  The SED of two-phase torus model has been computed with a self-consistent three dimensional radiative transfer code, with a  set of model parameters including the viewing angle, the inner radius, the volume filling factor, optical depth of the clouds and the optical depth of the disk mid-plane. 

 To represent the cold dust emission, we employ a gray body model, which is defined as:
\begin{equation}\label{equ:gb}
  S_{\lambda}\propto(1-e^{-(\frac{\lambda_0}{\lambda})^{\beta}}) B_\lambda(T_{\rm dust})
\end{equation}
where $B_\lambda$ is the Planck blackbody spectrum, $T_{\rm dust}$ is dust temperature. 
We adopt $\lambda_0$ = 125$\,\mu$m and $\beta$=1.6, which are the typical values  used for high redshift QSOs \citep{beelen2006,wang2008,wang2011}.
The dust temperature $T_{\rm dust}$ is set as a free parameter with a uniform prior truncated to the  interval of $log(T_{\rm dust}$/K)=[1,2].

Adopting a combination of the CLUMPY or two-phase torus  model and a gray body, we implement the IR SED decomposition of the ten Hot DOGs in our sample by using an updated version of the Bayesian SED fitting code, BayeSED \citep{han2012, han2014}\footnote{https://bitbucket.org/hanyk/bayesed/}. BayeSED is designed to be a general purpose Bayesian SED fitting code. It can be used to fit the multiwavelength SEDs of galaxies with a combination of whatever SED models. A detailed description of BayeSED can be found in \citet{fan2016b}. Here we summarize several basic features of BayeSED. (1)  By using principal component analysis (PCA), the SED library dimensionality can be reduced without sacrificing much accuracy. (2) The model SED at any position of the parameter space spanning the model SED library can be generated by employing a supervised machine learning method (K-Nearest Neighbors searching). (3) Instead of the traditional Markov Chain Monte Carlo (MCMC) algorithm, the multimodal nested sampling algorithm \citep{feroz2008,feroz2009}  has been employed  to obtain the posterior probability distribution function (PDF) of parameters. 

\section{Results}\label{sec:res}

\begin{figure}
\plotone{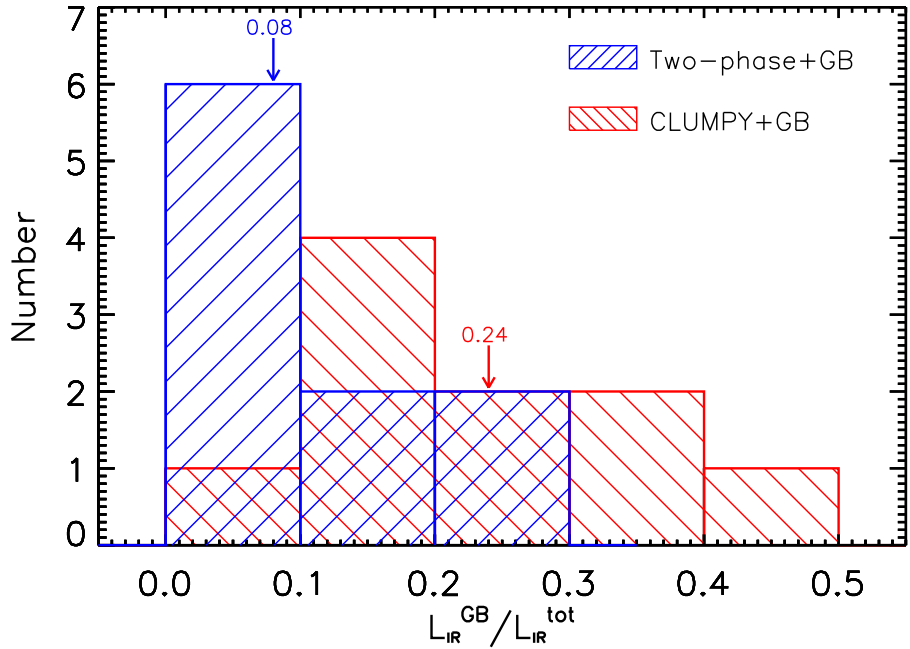}
\caption{Distributions of the fraction of the cold dust emission to the total IR luminosity ($L_{\rm IR}^{\rm GB}/L_{\rm IR}^{\rm tot}$). The mean values of $L_{\rm IR}^{\rm GB}/L_{\rm IR}^{\rm tot}$  are 0.24 and 0.08 in the CLUMPY+GB model (red) and the Two-phase+GB model (blue), respectively.}\label{fig:distlcold}
\end{figure}

\subsection{IR luminosities and dust temperatures}\label{subsec:lir}

\begin{table*}
\centering
\caption{Derived IR luminosities of Hot DOGs.}\label{tab:lum}
\begin{tabular}{ccccccc}
\toprule
              & \multicolumn{3}{c}{CLUMPY torus + Gray body} &  \multicolumn{3}{c}{Two-phase torus + Gray body} \\
\cmidrule(l){2-4}\cmidrule(l){5-7}
  Source      & log $L_{\rm IR}^{\rm t}$ & log $L_{\rm IR}^{\rm GB}$ &   log $L_{\rm IR}^{\rm tot}$  & log $L_{\rm IR}^{\rm t}$ & log $L_{\rm IR}^{\rm GB}$ &   log $L_{\rm IR}^{\rm tot}$   \\
              & ($L_\odot$)      & ($L_\odot$)       & ($L_\odot$)           & ($L_\odot$)      & ($L_\odot$)       & ($L_\odot$)            \\
\midrule
W0126$-$0529  &  13.99$\pm$0.01  &   13.92$\pm$0.01  &   14.26$\pm$0.03      &   14.20$\pm$0.01 &   -               &  14.20$\pm$0.01              \\
W0134$-$2922  &  14.02$\pm$0.02  &   13.20$\pm$0.05  &   14.08$\pm$0.08      &   14.07$\pm$0.17 &   12.68$\pm$0.48  &  14.10$\pm$0.65                      \\
W0248+2705    &  13.43$\pm$0.05  &   13.11$\pm$0.07  &   13.60$\pm$0.12      &   13.53$\pm$0.05 &   -               &  13.53$\pm$0.05                      \\
W0422$-$1028  &  13.45$\pm$0.03  &   13.01$\pm$0.05  &   13.58$\pm$0.08      &   13.52$\pm$0.09 &   12.64$\pm$0.34  &  13.58$\pm$0.43                      \\
W0757+5113    &  13.42$\pm$0.03  &   12.80$\pm$0.05  &   13.51$\pm$0.08      &   13.51$\pm$0.08 &   12.80$\pm$0.17  &  13.59$\pm$0.25                      \\
W0859+4823    &  13.99$\pm$0.02  &   13.36$\pm$0.03  &   14.08$\pm$0.05      &   14.04$\pm$0.06 &   -               &  14.04$\pm$0.06                      \\
W1248$-$2154  &  14.13$\pm$0.02  &   13.36$\pm$0.07  &   14.20$\pm$0.09      &   14.25$\pm$0.11 &   13.21$\pm$0.35  &  14.32$\pm$0.46                      \\
W1838+3429    &  14.16$\pm$0.02  &   -               &   14.16$\pm$0.02      &   14.24$\pm$0.19 &   -               &  14.24$\pm$0.19                      \\
W2201+0226    &  13.92$\pm$0.03  &   13.72$\pm$0.02  &   14.14$\pm$0.05      &   13.97$\pm$0.11 &   13.44$\pm$0.17  &  14.09$\pm$0.28                      \\
W2210$-$3507  &  13.97$\pm$0.02  &   13.50$\pm$0.02  &   14.10$\pm$0.04      &   14.04$\pm$0.04 &   13.52$\pm$0.03  &  14.16$\pm$0.07                      \\
\bottomrule
\end{tabular}
\end{table*}

In Figure \ref{fig:fitn08}, we present the best-fitting results with the CLUMPY+GB model for the ten SCUBA-2 observed Hot DOGs in our sample. Based on the best-fitting results of our IR SED decomposition approach, we can derive not only the total IR luminosities, but also the individual contributions of both torus (blue dashed line in Figure \ref{fig:fitn08}) and cold dust components (green dot-dashed line in Figure \ref{fig:fitn08}).  In Table \ref{tab:lum}, we list the total IR luminosities ($L_{\rm IR}^{\rm tot}$),  the IR luminosities emitted by the torus ($L_{\rm IR}^{\rm t}$) and cold dust ($L_{\rm IR}^{\rm GB}$), respectively. All of them are HyLIRGs (Hyper-luminous Infrared Galaxies) with $L_{\rm IR}^{\rm tot}>10^{13} L_\odot$. Seven out of ten Hot DOGs are ELIRGs (Extremely Luminous Infrared Galaxies, Tsai et al. 2015) with $L_{\rm IR}^{\rm tot}>10^{14} L_\odot$. The torus emissions contribute over half of the total IR energy output. However, all but one (W1838+3429) show the non-negligible IR luminosities ($10^{12.8-13.9} L_\odot$) emitted by cold dust, which contribute up to 50\% of the total IR luminosities.  The average fraction of the cold dust to the total IR luminosities is  24\% (see  Figure \ref{fig:distlcold}).  The results suggest that Hot DOGs contain both very powerful AGN and extreme starbursts with SFRs of several thousands  solar masses per year. 

\begin{figure}
\centering
\includegraphics[scale=0.9]{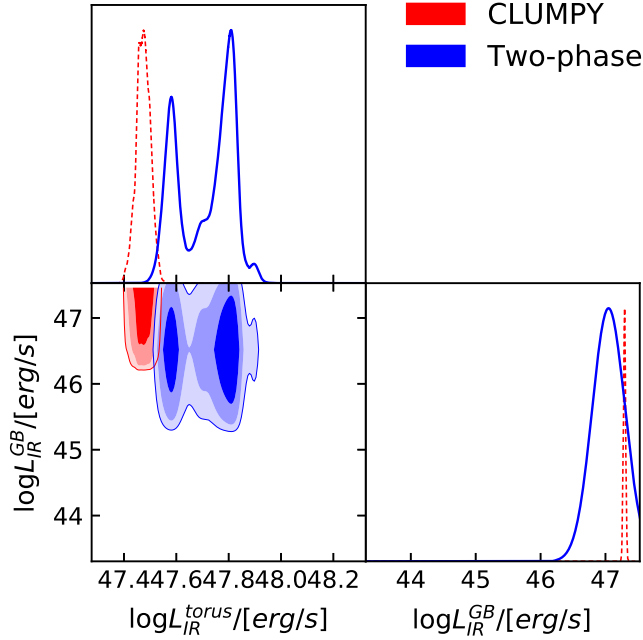}
\caption{One-dimensional and two-dimensional marginalized posterior probability distributions of IR luminosities of the torus ($L_{\rm IR}^{\rm t}$) and cold dust ($L_{\rm IR}^{\rm GB}$) components, for W2201+0226 as an example. The color coding represents the confidence levels. Both one-dimensional and two-dimensional marginalized posterior probability distributions have been normalized to unit area. Red and blue colors mark the CLUMPY+GB model and the Two-phase+GB model, respectively.}\label{fig:ppd}
\end{figure}

The derived IR luminosities may depend on the adopted torus models. In Figure \ref{fig:fits15}, we also present the best-fitting results with the Two-phase+GB model for the ten SCUBA-2 observed Hot DOGs in our sample.  The total IR luminosities ($L_{\rm IR}^{\rm tot}$),  the IR luminosities emitted by the torus ($L_{\rm IR}^{\rm t}$) and cold dust ($L_{\rm IR}^{\rm GB}$) are also listed in Table \ref{tab:lum}. We can find that the derived $L_{\rm IR}^{\rm tot}$ are independent on the adopted models. The average difference of the total IR luminosities between the CLUMPY+GB model and the Two-phase+GB model is only 0.06 dex.  However, the relative contributions of $L_{\rm IR}^{\rm t}$ and $L_{\rm IR}^{\rm GB}$ to $L_{\rm IR}^{\rm tot}$ change a bit.  Four out of ten Hot DOGs can be fitted with a sole two-phase torus model, while the rest of them require a cold dust component, contributing up to 30\% of the total IR luminosities. The average fraction of the cold dust to the total IR luminosities is  8\%, which is only one third of the fraction with the CLUMPY+GB model  (see  Figure \ref{fig:distlcold}).  In Figure \ref{fig:ppd}, we plot the one-dimensional and two-dimensional marginalized posterior probability distributions of IR luminosities of the torus ($L_{\rm IR}^{\rm t}$) and cold dust ($L_{\rm IR}^{\rm GB}$) components, for W2201+0226 as an example. We can find that both $L_{\rm IR}^{\rm t}$ and $L_{\rm IR}^{\rm GB}$ have been constrained to a narrower range using the CLUMPY+GB model than using the Two-phase+GB model. 

We employ a  gray body model to represent the cold dust emission, which is defined in Equation \ref{equ:gb} by assuming general opacity. The cold dust temperature $T_{\rm dust}$ of each Hot DOG has been derived if a gray body component is required. In the CLUMPY+GB model, the derived dust temperatures range from 42 K to 91 K, with a mean value of 68 K.  As a comparison, the derived dust temperatures in the Two-phase+GB model range from 36 K to 74 K, with a mean value of 56 K. 

One of the advantages of BayeSED code is that we can compare different models quantitatively by deriving their Bayesian evidences \citep{JeffreysH1998a,Gregory2005a}, which represent a practical implementation of the Occam's razor principle.  In Table \ref{tbl-ev}, we present the natural logarithm ${\rm ln(ev_{CLUMPY+GB})}$ and ${\rm ln(ev_{Two-phase+GB}})$ of the Bayesian evidences for CLUMPY+GB and Two-phase+GB models. We also present the natural logarithm of Bayes factor ${\rm ln} (\frac{{\rm ev_{CLUMPY+GB}}}{{\rm ev_{Two-phase+GB}}}$) in Table \ref{tbl-ev}. We find that the CLUMPY+GB model has the higher Bayesian evidence than the Two-phase+GB model for all Hot DOGs. We also find that eight out of ten Hot DOGs have ${\rm ln} (\frac{{\rm ev_{CLUMPY+GB}}}{{\rm ev_{Two-phase+GB}}}) > 10$ (corresponding to odds of $> 20000:1$), which represent strong evidence in favor of CLUMPY+GB model according to the empirically calibrated Jeffreys's scale \citep{jeffreys1961,trotta2008}. This result indicates that the cold dust component in Hot DOGs may be non-negligible. 

\begin{table}
\centering
\caption{The Bayesian evidences of "CLUMPY+GB" and "Two-phase+GB" models\label{tbl-ev}}
\begin{tabular}{lccc}
\hline
\hline
Source & ln(ev$_{\rm CLUMPY+GB}$) & ln(ev$_{\rm Two-phase+GB}$) & ln ($\frac{{\rm ev_{CLUMPY+GB}}}{{\rm ev_{Two-phase+GB}}}$) \\
\hline

     W0126$-$0529  &  $-145.56\pm0.32$  &  $-210.03\pm0.27$  &  $64.47\pm0.58$	\\
     W0134$-$2922  &  $ -77.97\pm0.20$  &  $-100.19\pm0.20$  &  $22.23\pm0.40$	\\
     W0248+2705    &  $ -71.88\pm0.17$  &  $ -87.45\pm0.18$  &  $15.57\pm0.35$	\\
     W0422$-$1028  &  $ -71.84\pm0.18$  &  $ -93.43\pm0.20$  &  $21.59\pm0.38$	\\
     W0757+5113    &  $ -69.94\pm0.17$  &  $ -84.46\pm0.24$  &  $14.52\pm0.42$	\\
     W0859+4823    &  $ -94.29\pm0.19$  &  $-112.32\pm0.20$  &  $18.03\pm0.40$	\\
     W1248$-$2154  &  $ -69.40\pm0.18$  &  $ -76.50\pm0.24$  &  $ 7.10\pm0.42$	\\
     W1838+3429    &  $ -41.46\pm0.17$  &  $ -44.85\pm0.20$  &  $ 3.39\pm0.38$	\\
     W2201+0226    &  $ -87.25\pm0.21$  &  $-104.50\pm0.22$  &  $17.25\pm0.43$	\\
     W2210$-$3507  &  $ -99.27\pm0.20$  &  $-118.57\pm0.26$  &  $19.30\pm0.46$	\\
  
\hline
\end{tabular}
\end{table}

\subsection{Environments around Hot DOGs}

\begin{figure*}
\plotone{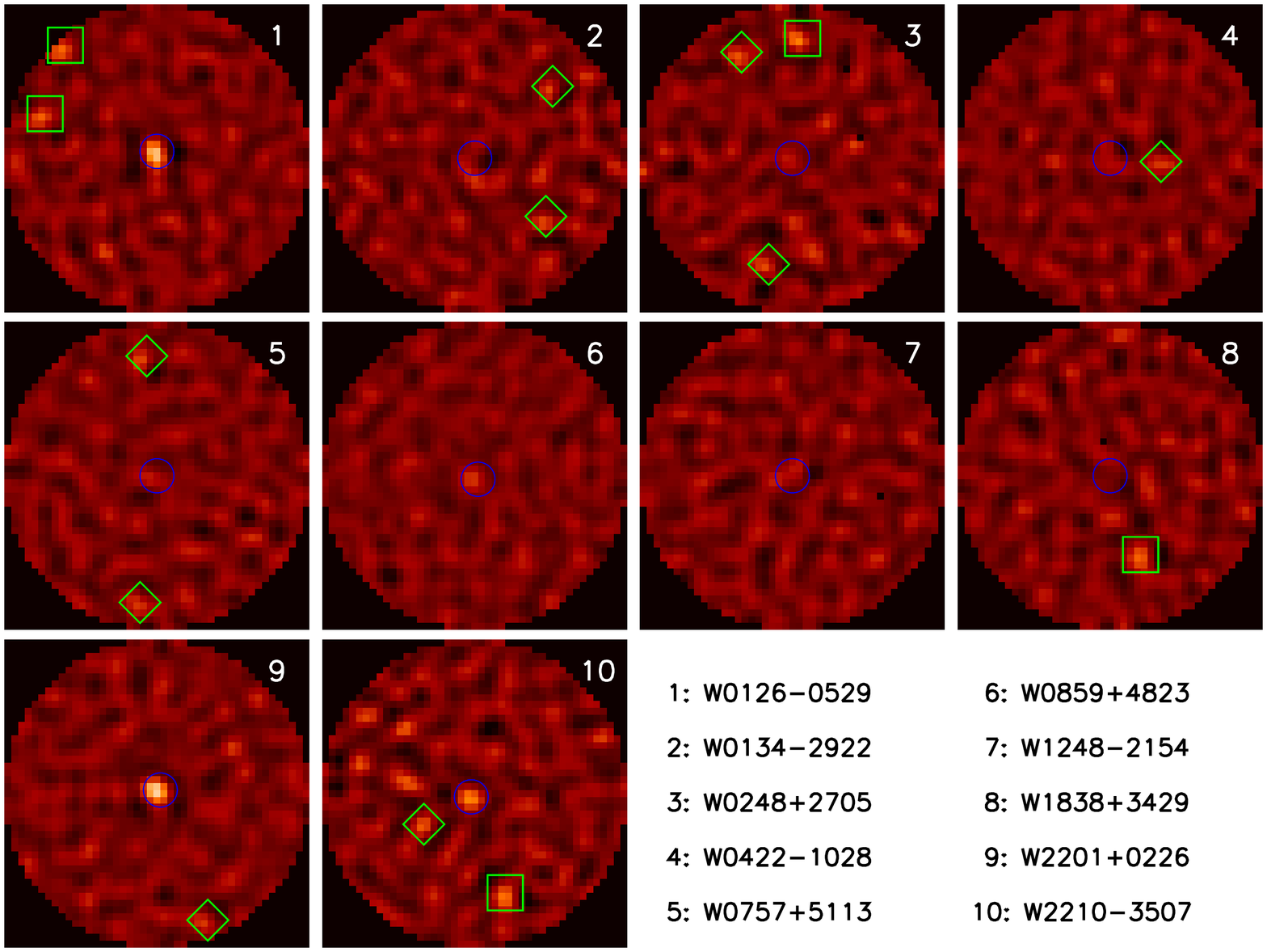}
\caption{SCUBA-2 850$\mu m~$ 1.5-arcminute-radius maps of the ten Hot DOGs in our sample. The circles mark the position of the WISE detections of Hot DOGs.  Serendipitous sources brighter than 3$\sigma$ and 3.5$\sigma$  are shown by diamonds and squares, respectively. In total, 14 serendipitous sources at $>3\sigma$ have been detected, while the number falls to 5 at $>3.5\sigma$ level. Having a good sensitivity, \citet{jones2014} found 17 and 7 serendipitous sources at $>3\sigma$ and $>3.5\sigma$ around 10 Hot DOGs, respectively.  }\label{fig:scuba2map}
\end{figure*}

Hot DOGs have been found to live in dense environments by previous works \citep{jones2014,assef2015,jones2015,silva2015}. Fourteen and five submillimeter serendipitous sources in the 10 SCUBA-2 fields around Hot DOGs have been detected at greater than 3$\sigma$ and 3.5$\sigma$, respectively (see Figure \ref{fig:scuba2map} and Table \ref{tab:number}). As indicated by the Equation 7 in \citet{geach2017}, the false detection rate of SCUBA-2 850 $\mu m$ observations is a function of signal-to-noise ratio. The false detection rate is about 60\% at the $>3\sigma$ level and falls to about 20\% at $3.5\sigma$ limit. We calculate the cumulative source counts N($>{\rm S_{850 \mu m}}$) for the submillimeter serendipitous sources with a S/N $>3.5$  by following the method described in \citet{ono2014} and \citet{silva2015}. We estimate the deboosted flux densities by correcting the flux boosting using the Equation 5 in \citet{geach2017}. 

\begin{figure}
\plotone{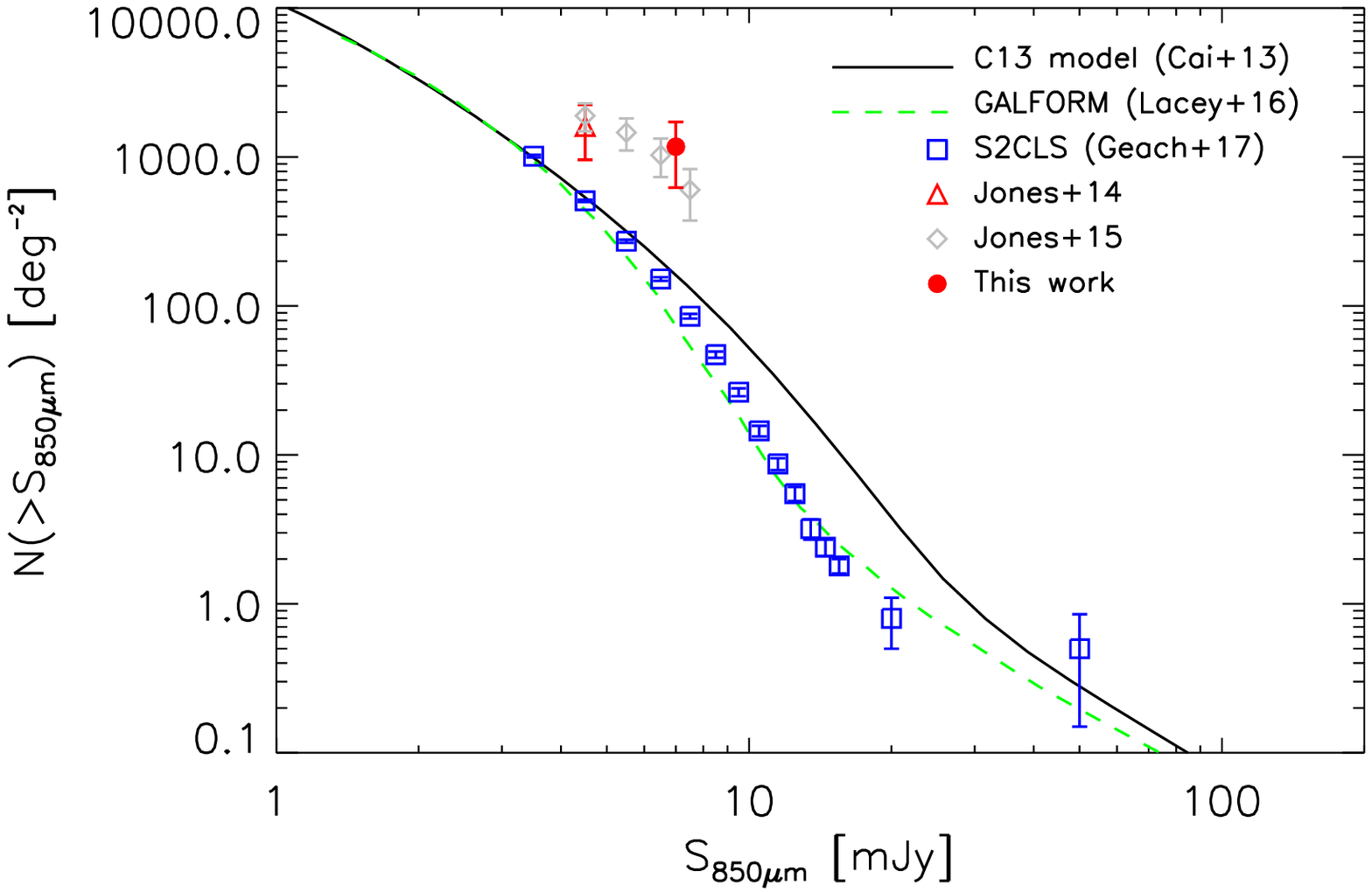}
\caption{The cumulative number count of SMGs around 10 Hot DOGs (red circle) obtained with SCUBA-2 observations at 850 $\mu m$ with a S/N $>3.5$.  We also plot the SCUBA-2 counts around 30 radio-selected Hot DOGs (gray diamonds) from \citet{jones2015} and 10 WISE-selected Hot DOGs (red triangle) from \citet{jones2014}. We overplot the number counts of SCUBA-2 850 $\mu m$ sources determined by \citet{geach2017} in the S2CLS survey (blue squares) and the models of \cite{cai2013} and \citet{lacey2016}.}\label{fig:numb}
\end{figure}

In Figure \ref{fig:numb}, we plot our result (red circle) and compare it with those in the previous results around Hot DOGs, the blank field survey and expected from models. We include the counts  around 30 radio-selected Hot DOGs (gray diamonds) from \citet{jones2015} and 10 WISE-selected Hot DOGs (red triangle) from \citet{jones2014}. Both samples used to estimate number counts have been selected from \citet{jones2017} with S/N $>3.5$. We overplot the number counts of SCUBA-2 850 $\mu m$ sources determined by \citet{geach2017} in SCUBA-2 Cosmology Legacy Survey (S2CLS, blue squares) and the models of \cite{cai2013} and \citet{lacey2016}. Considering the uncertainties, our count is consistent with the counts obtained by \citet{jones2015}. Relative to blank fields and model expectations, the number counts of SMGs around Hot DOGs imply an excess of $\sim 6$ times. 

\begin{figure*}
\plotone{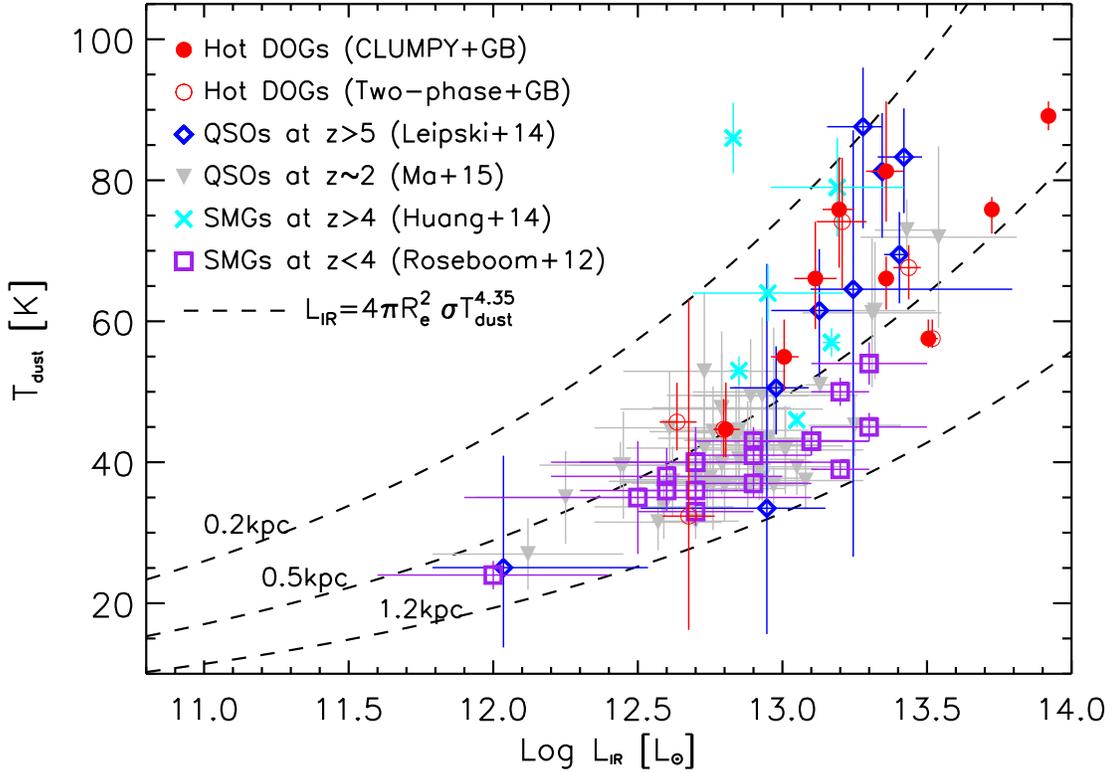}
\caption{Temperature ($T_{\rm dust}$) versus IR luminosity ($L_{\rm IR}^{\rm GB}$) of cold dust  in Hot DOGs derived by our IR SED decomposition with CLUMPY+GB model (filled circles) and Two-phase+GB model (open circles), respectively.  We also plot the results of other high-redshift populations: QSOs at $z>5$ (diamonds; Leipski et al. 2014) and at $z\sim2$ (upside down triangles; Ma \& Yan 2015), SMGs at $z>4$ (crosses; Huang et al. 2014) and at $z<4$ (squares; Roseboom et al. 2012). Dashed lines represent the expected $T_{\rm dust}-L_{\rm IR}^{\rm GB}$ relation of a gray body with different effective radius values (0.2, 0.5 and 1.2 kpc). The future high-resolution dust continuum observations by ALMA will provide further insights. }\label{fig:tdustlir}
\end{figure*}

\section{Discussion}\label{sec:dis}

\begin{table}
\centering
\caption{Number of serendipitous sources and negative peaks in each map at greater than 3$\sigma$ and 3.5$\sigma$, respectively.}\label{tab:number}
\begin{tabular}{ccccc}
\toprule
\multirow{3}{*}{Source}    & \multicolumn{2}{c}{Number of } &  \multicolumn{2}{c}{Number of} \\
                           & \multicolumn{2}{c}{serendipitous sources} &  \multicolumn{2}{c}{negative peaks} \\
\cmidrule(l){2-3}\cmidrule(l){4-5}
              &  $>3\sigma$     &   $>3.5\sigma$    &  $>3\sigma$     &   $>3.5\sigma$     \\
\midrule
W0126$-$0529  &     2           &       2           &      1          &    1                 \\
W0134$-$2922  &     2           &       0           &      0          &    0           \\
W0248+2705    &     3           &       1           &      3          &    0           \\
W0422$-$1028  &     1           &       0           &      1          &    0           \\
W0757+5113    &     2           &       0           &      3          &    0           \\
W0859+4823    &     0           &       0           &      1          &    1           \\
W1248$-$2154  &     0           &       0           &      0          &    0           \\
W1838+3429    &     1           &       1           &      0          &    0           \\
W2201+0226    &     1           &       0           &      0          &    0           \\
W2210$-$3507  &     2           &       1           &      0          &    0           \\
\bottomrule
\end{tabular}
\end{table}

Combined with other IR data, our SCUBA-2 follow-up observations of ten Hot DOGs show that they have very high IR luminosities and cold dust temperatures. Hot DOGs are brighter by an order of magnitude than other IR luminous galaxies, such as SMGs and DOGs,  but they have the comparable bolometric luminosities to the most luminous QSOs in the Universe \citep{assef2015}. Our IR SED decomposition of Hot DOGs suggests that most of their IR radiations are from AGN tori, which are  heated by the black-hole accretion. This result confirms again that Hot DOGs are in fact the luminous, dust-obscured QSOs.  Despite the dominance of AGN torus component, most of Hot DOGs still require a non-negligible cold dust component, suggesting a co-existing starburst activity. Our recent ALMA CO observations of three Hot DOGs reveal that they have a significant molecular gas reservoir ($\sim10^{10-11} M_\odot$) which can provide the required fuel for both extreme star formation and black-hole accretion (Fan et al. 2017, in prep). 

The cold dust temperature $T_{\rm dust}$ depends on the adopted torus and gray body models in the IR SED decomposition. The choice of torus model will affect both the derived temperature and IR luminosity of cold dust. Adopting CLUMPY torus model, we derive the higher values of $T_{\rm dust}$ and $L_{\rm IR}^{\rm GB}$  (see Table \ref{tab:lum} and Figure \ref{fig:tdustlir}).  For six Hot DOGs having the SED decomposition results with both torus models, the derived $log L_{\rm IR}^{\rm GB}$ and $T_{dust}$ using the CLUMPY+GB model  are on average 0.2 dex and 12 K higher than those derived using the Two-phase+GB model, respectively. The difference may raise from the truth that two torus models adopt the different dust particles and torus geometry. The  starburst contribution to the far IR emission can well be marginal in the Two-phase+GB model, which is consistent with the result in \citet{siebenmorgen2015}.  In Equation \ref{equ:gb}, we assume general opacity. As discussed in our previous work \citep{fan2016b}, the derived cold dust temperature of Hot DOGs adopting general opacity is averagely $\sim 20$ K higher than that under the optically thin assumption.  This trend also exists in other high-redshift IR luminous galaxies \citep{conley2011,magdis2012}. However, the derived cold dust temperature of our Hot DOGs is about 56 K adopting the optically thin assumption, which is still higher than those found in other high-redshift IR luminous galaxies \citep{magdis2010, magnelli2012,melbourne2012}. 

In Figure \ref{fig:tdustlir}, we plot the derived dust temperature as a function of IR luminosity for our Hot DOGs. We also plot the $T_{\rm dust}-L_{\rm IR}$ relation of other high-redshift populations as a comparison, including QSOs at $z>5$ (diamonds; Leipski et al. 2014) and at $z\sim2$ (upside down triangles; Ma \& Yan 2015), SMGs at $z>4$ (crosses; Huang et al. 2014) and at $z<4$ (squares; Roseboom et al. 2012). We use a modified Stefan-Boltzmann  law ($L_{\rm IR}=4\pi R_e^2\sigma T^{4.35}$) to describe $T_{\rm dust}-L_{\rm IR}$ relation of a gray body \citep{symeonidis2013,ma2015,fan2016b}.  As seen in Figure \ref{fig:tdustlir}, Hot DOGs have hotter dust temperature than SMGs. The higher dust temperature in our Hot DOGs may be due to two aspects: their high IR luminosity $L_{\rm IR}$ and/or their small effective radius of the equivalent FIR-emitting region ($R_e$), as suggested by the expected $T_{\rm dust}-L_{\rm IR}$ relation of a gray body (dashed lines in Figure \ref{fig:tdustlir}). The future ALMA high-resolution dust continuum observations will provide further insights by comparing the FIR-emitting sizes of Hot DOGs and SMGs. Compared to UV-selected QSOs with similar IR luminosity, Hot DOGs have the same dust temperature range, confirming the idea that Hot DOGs are the luminous dust-obscured QSOs. 

By counting the submillimeter serendipitous sources around Hot DOGs and estimating their cumulative number counts, we find that Hot DOGs live in dense environments (see Figure \ref{fig:scuba2map},\ref{fig:numb}). The result is consistent with the previous works.  \citet{jones2014} found evidence of an overdensity of sub-mm neighbors to a small sample of HotDOGs using 850 $\mu m$ observations with SCUBA-2 at JCMT. \citet{assef2015} studied the environments of HotDOGs using the deep Warm\ Spitzer IRAC imaging. They showed that the number of red galaxies within 1 arcminute radius is significantly above the number observed in random pointing. And their environments are as dense as those of the clusters identified by the Clusters Around Radio-loud AGN survey (CARLA). Using the Atacama Large Millimeter/Sub-millimeter Array (ALMA) 345 GHz images, \citet{silva2015} also found an overdensity of SMGs around radio-selected, Hot DOGs at $z\sim2$. The results are not surprising, given the extremely high luminosity and stellar mass ($>10^{11} M_\odot$) of HotDOGs \citep{assef2015,lonsdale2015}. \citet{frey2016} performed exploratory high-resolution 1.7-GHz observations of a small sample of four Hot DOGs  with the European VLBI Network (EVN). The high-resolution radio structures, the component separations and radio powers support that QSOs residing in Hot DOGs may be genuine young radio sources where starburst and QSO activities coexist. This result leads to the speculation that Hot DOGs could be the earlier evolutionary phase of radio-loud QSOs, which are preferentially located in high-redshift protoclusters \citep{wylezalek2013,hatch2014}. In order to study the environments of HotDOGs thoroughly, we are proposing to search Lyman alpha emitters (LAEs) around them. As the narrow-band imaging can constrain the redshift of the objects to the small range, LAEs are very useful in studying the spatial distribution of the star-forming galaxies at high redshift. In the recent decade, targeted searches using narrow-band imaging for high-redshift rich environments around powerful high-redshift radio galaxies (HzRGs) have proven very successful \citep{venemans2007,matsuda2011}.  Our ongoing studies will not only provide a direct evidence for (or against) whether Hot DOGs could be a good tracer of high-redshift galaxy structure, but also provide an insight to whether their extreme starburst and AGN activities are related to their environments.

\section{Summary}\label{sec:sum}

Our main results from JCMT SCUBA-2 850 $\mu m$ follow-up of 10 Hot DOGs are summarized as follows.

\begin{enumerate}
\item The IR SED of Hot DOGs in our sample can be fitted by a combination of AGN torus and cold dust components. The derived total IR luminosities of Hot DOGs are extremely high and independent of the choice of torus model. Most of them are ELIRGs with $L_{\rm IR}^{\rm tot}>10^{14} L_\odot$.
\item The torus emissions dominate the total IR energy output, regardless of the choice of torus model. The derived cold dust contribution depends on the adopted torus model.  If adopting the Two-phase+GB model, the average fraction of the cold dust to the total IR luminosity is  about 8\%, which is only one third of the fraction with the CLUMPY+GB model. We compare two models quantitatively by deriving their Bayesian evidences. We find that the CLUMPY+GB model has the significantly higher Bayesian evidence than the Two-phase+GB model for all Hot DOGs, which represent strong evidence in favor of CLUMPY+GB model. This result indicates that the cold dust component in Hot DOGs may be non-negligible. 
\item The derived cold dust temperatures in Hot DOGs are on average 68 K (56 K)  using the CLUMPY+GB model (Two-phase+GB model), which are comparable to those in QSOs with similar IR luminosity, but much higher than those in SMGs. Higher dust temperatures in Hot DOGs may be due to the more intense radiation field caused by intense starburst and obscured AGN activities. 
\item  We count the submillimeter serendipitous sources around Hot DOGs and estimate their cumulative number counts at $>3.5\sigma$ level. The number counts of SMGs around Hot DOGs imply an excess of $\sim 6$ times relative to blank fields and model expectations. The result confirms the previous argument that Hot DOGs may lie in dense environments. Future studies with narrow-band imaging may increase our insight on their environments.
\item All of our results support the scenario in which Hot DOGs are luminous, dust-obscured QSOs, linking extreme starburst and UV-bright QSOs.
\end{enumerate}

\begin{acknowledgements}
The authors would like to thank the anonymous referee for his/her comments and suggestions, which have greatly improved this paper. LF acknowledges the support from National Key Research and Development Program of China (No. 2017YFA0402703). This work is supported by the National Natural Science Foundation of China (NSFC, Nos. 11433005, 11773063 and 11773020) and Shandong Provincial Natural Science Foundation, China (ZR2017QA001). KK acknowledges the Knut and Alice Wallenberg Foundation for support. 

The James Clerk Maxwell Telescope is operated by the East Asian Observatory on behalf of The National Astronomical Observatory of Japan, Academia Sinica Institute of Astronomy and Astrophysics, the Korea Astronomy and Space Science Institute, the National Astronomical Observatories of China and the Chinese Academy of Sciences (Grant No. XDB09000000), with additional funding support from the Science and Technology Facilities Council of the United Kingdom and participating universities in the United Kingdom and Canada.

\facility{\textit{WISE}, \textit{Herschel} (PACS,SPIRE), \textit{JCMT} (SCUBA-2)}.

\end{acknowledgements}




\begin{thebibliography}{99}

\bibitem[Alexander \& Hickox(2012)]{alexander2012} Alexander, D.~M., \& Hickox, R.~C.\ 2012, New Astronomy Reviews , 56, 93
\bibitem[Assef et al.(2015)]{assef2015} Assef, R.~J., Eisenhardt, P.~R.~M., Stern, D., et al.\ 2015, \apj, 804, 27
\bibitem[Assef et al.(2016)]{assef2016} Assef, R.~J., Walton, D.~J., Brightman, M., et al.\ 2016, \apj, 819, 111
\bibitem[Beelen et al.(2006)]{beelen2006} Beelen, A., Cox, P., Benford, D.~J., et al.\ 2006, \apj, 642, 694
\bibitem[Blain et al. (2002)]{blain2002} Blain, A.~W., Smail, I., Ivison, R.~J., Kneib, J.-P., \& Frayer, D.~T.\ 2002, \physrep, 369, 111
\bibitem[Brandt \& Alexander(2015)]{brandt2015} Brandt, W.~N., \& Alexander, D.~M.\ 2015, \aapr, 23, 1
\bibitem[Bridge et al.(2013)]{bridge2013} Bridge, C.~R., Blain, A., Borys, C.~J.~K., et al.\ 2013, \apj, 769, 91
\bibitem[Cai et al.(2013)]{cai2013} Cai, Z.-Y., Lapi, A., Xia, J.-Q., et al.\ 2013, \apj, 768, 21 
\bibitem[Casey et al.(2014)]{casey2014} Casey, C.~M., Narayanan, D., \& Cooray, A.\ 2014, \physrep, 541, 45
\bibitem[Chapin et al.(2013)]{chapin2013} Chapin, E.~L., Berry, D.~S., Gibb, A.~G., et al.\ 2013, \mnras, 430, 2545
\bibitem[Conley et al.(2011)]{conley2011} Conley, A., Cooray, A., Vieira, J.~D., et al.\ 2011, \apjl, 732, L35
\bibitem[Cutri et al.(2013)]{cutri2013} Cutri, R.~M., \& et al.\ 2013, VizieR Online Data Catalog, 2328, 0
\bibitem[Dempsey et al.(2013)]{dempsey2013} Dempsey, J.~T., Friberg, P., Jenness, T., et al.\ 2013, \mnras, 430, 2534
\bibitem[Dey et al.(2008)]{dey2008} Dey, A., Soifer, B.~T., Desai, V., et al.\ 2008, \apj, 677, 943
\bibitem[D{\'{\i}}az-Santos et al.(2016)]{diaz-santos2016} D{\'{\i}}az-Santos, T., Assef, R.~J., Blain, A.~W., et al.\ 2016, \apjl, 816, L6
\bibitem[Eisenhardt et al.(2012)]{eisenhardt2012} Eisenhardt, P.~R.~M., Wu, J., Tsai, C.-W., et al.\ 2012, \apj, 755, 173
\bibitem[Fan et al.(2016a)]{fan2016a} Fan, L., Han, Y., Fang, G., et al.\ 2016, \apjl, 822, L32 
\bibitem[Fan et al.(2016b)]{fan2016b} Fan, L., Han, Y., Nikutta, R., Drouart, G., \& Knudsen, K.~K.\ 2016, \apj, 823, 107 
\bibitem[Farrah et al.(2017)]{farrah2017} Farrah, D., Petty, S., Connolly, B., et al.\ 2017, arXiv:1705.02649 
\bibitem[MultiNest, Feroz \& Hobson(2008)]{feroz2008} Feroz, F., \& Hobson, M.~P.\ 2008, \mnras, 384, 449
\bibitem[Feroz et al.(2009)]{feroz2009} Feroz, F., Hobson, M.~P., \& Bridges, M.\ 2009, \mnras, 398, 1601
\bibitem[Frey et al.(2016)]{frey2016} Frey, S., Paragi, Z., Gab{\'a}nyi, K.~{\'E}., \& An, T.\ 2016, \mnras, 455, 2058
\bibitem[Geach et al.(2017)]{geach2017} Geach, J.~E., Dunlop, J.~S., Halpern, M., et al.\ 2017, \mnras, 465, 1789 
\bibitem[{Gregory(2005)}]{Gregory2005a} Gregory, P. 2005, {B}ayesian {L}ogical {D}ata {A}nalysis for the {P}hysical {S}ciences (New York, NY, USA: Cambridge University Press)
\bibitem[Griffin et al.(2010)]{griffin2010} Griffin, M.~J., Abergel, A., Abreu, A., et al.\ 2010, \aap, 518, L3
\bibitem[{{Han} \& {Han}(2012)}]{han2012} {Han}, Y., \& {Han}, Z. 2012, \apj, 749, 123
\bibitem[{{Han} \& {Han}(2014)}]{han2014} {Han}, Y., \& {Han}, Z. 2014, \apjs, 215, 2
\bibitem[Holland et al.(2013)]{holland2013} Holland, W.~S., Bintley, D., Chapin, E.~L., et al.\ 2013, \mnras, 430, 2513
\bibitem[Hopkins et al.(2008)]{hopkins2008} Hopkins, P.~F., Hernquist, L., Cox, T.~J., \& Kere{\v s}, D.\ 2008, \apjs, 175, 356
\bibitem[Hatch et al.(2014)]{hatch2014} Hatch, N.~A., Wylezalek, D., Kurk, J.~D., et al.\ 2014, \mnras, 445, 280
\bibitem[Huang et al.(2014)]{huang2014} Huang, J.-S., Rigopoulou, D., Magdis, G., et al.\ 2014, \apj, 784, 52
\bibitem[Jeffreys (1961)]{jeffreys1961} Jeffreys, H. 1961, The Theory of Probability Oxford Classics Series (3rd ed.; Oxford: Oxford Univ. Press)
\bibitem[{Jeffreys(1998)}]{JeffreysH1998a} Jeffreys, H. 1998, The Theory of Probability (OUP Oxford)
\bibitem[Jones et al.(2014)]{jones2014} Jones, S.~F., Blain, A.~W., Stern, D., et al.\ 2014, \mnras, 443, 146
\bibitem[Jones et al.(2015)]{jones2015} Jones, S.~F., Blain, A.~W., Lonsdale, C., et al.\ 2015, \mnras, 448, 3325
\bibitem[Jones et al.(2017)]{jones2017} Jones, S.~F., Blain, A.~W., Assef, R.~J., et al.\ 2017, \mnras, 469, 4565 
\bibitem[Komatsu et al.(2011)]{komatsu2011} Komatsu, E., Smith, K.~M., Dunkley, J., et al.\ 2011, \apjs, 192, 18
\bibitem[Kormendy \& Ho(2013)]{kormendy2013} Kormendy, J., \& Ho, L.~C.\ 2013, \araa, 51, 511
\bibitem[Lacey et al.(2016)]{lacey2016} Lacey, C.~G., Baugh, C.~M., Frenk, C.~S., et al.\ 2016, \mnras, 462, 3854
\bibitem[Leipski et al.(2014)]{leipski2014} Leipski, C., Meisenheimer, K., Walter, F., et al.\ 2014, \apj, 785, 154
\bibitem[Lonsdale et al.(2015)]{lonsdale2015} Lonsdale, C.~J., Lacy, M., Kimball, A.~E., et al.\ 2015, \apj, 813, 45 
\bibitem[Ma \& Yan(2015)]{ma2015} Ma, Z., \& Yan, H.\ 2015, \apj, 811, 58
\bibitem[Magdis et al.(2012)]{magdis2012} Magdis, G.~E., Daddi, E., B{\'e}thermin, M., et al.\ 2012, \apj, 760, 6
\bibitem[Magdis et al.(2010)]{magdis2010} Magdis, G.~E., Elbaz, D., Hwang, H.~S., et al.\ 2010, \mnras, 409, 22
\bibitem[Magnelli et al.(2012)]{magnelli2012} Magnelli, B., Lutz, D., Santini, P., et al.\ 2012, \aap, 539, A155
\bibitem[Matsuda et al.(2011)]{matsuda2011} Matsuda, Y., Smail, I., Geach, J.~E., et al.\ 2011, \mnras, 416, 2041
\bibitem[Melbourne et al.(2012)]{melbourne2012} Melbourne, J., Soifer, B.~T., Desai, V., et al.\ 2012, \aj, 143, 125 
\bibitem[Nenkova et al.(2002)]{nenkova2002} Nenkova, M., Ivezi{\'c}, {\v Z}., \& Elitzur, M.\ 2002, \apjl, 570, L9
\bibitem[{{Nenkova} {et~al.}(2008{\natexlab{a}}){Nenkova}, {Sirocky}, {Ivezi{\'c}}, \& {Elitzur}}]{nenkova2008a} {Nenkova}, M., {Sirocky}, M.~M., {Ivezi{\'c}}, {\v Z}., \& {Elitzur}, M.   2008{\natexlab{a}}, \apj, 685, 147
\bibitem[{{Nenkova} {et~al.}(2008{\natexlab{b}}){Nenkova}, {Sirocky}, {Nikutta}, {Ivezi{\'c}}, \& {Elitzur}}]{nenkova2008b} {Nenkova}, M., {Sirocky}, M.~M., {Nikutta}, R., {Ivezi{\'c}}, {\v Z}., \& {Elitzur}, M. 2008{\natexlab{b}}, \apj, 685, 160
\bibitem[Ono et al.(2014)]{ono2014} Ono, Y., Ouchi, M., Kurono, Y., \& Momose, R.\ 2014, \apj, 795, 5 
\bibitem[Piconcelli et al.(2015)]{piconcelli2015} Piconcelli, E., Vignali, C., Bianchi, S., et al.\ 2015, \aap, 574, L9
\bibitem[Pilbratt et al.(2010)]{pilbratt2010} Pilbratt, G.~L., Riedinger, J.~R., Passvogel, T., et al.\ 2010, \aap, 518, L1
\bibitem[Poglitsch et al.(2010)]{poglitsch2010} Poglitsch, A., Waelkens, C., Geis, N., et al.\ 2010, \aap, 518, L2
\bibitem[Reddy et al.(2008)]{reddy2008} Reddy, N.~A., Steidel, C.~C., Pettini, M., et al.\ 2008, \apjs, 175, 48
\bibitem[Ricci et al.(2017)]{ricci2017} Ricci, C., Assef, R.~J., Stern, D., et al.\ 2017, \apj, 835, 105 
\bibitem[Roseboom et al.(2012)]{roseboom2012} Roseboom, I.~G., Ivison, R.~J., Greve, T.~R., et al.\ 2012, \mnras, 419, 2758
\bibitem[Sanders et al.(1988)]{sanders1988} Sanders, D.~B., Soifer, B.~T., Elias, J.~H., et al.\ 1988, \apj, 325, 74
\bibitem[Sanders \& Mirabel (1996)]{sanders1996} Sanders, D.~B., \& Mirabel, I.~F.\ 1996, \araa, 34, 749
\bibitem[Siebenmorgen et al.(2015)]{siebenmorgen2015} Siebenmorgen, R., Heymann, F., \& Efstathiou, A.\ 2015, \aap, 583, A120
\bibitem[Silva et al.(2015)]{silva2015} Silva, A., Sajina, A., Lonsdale, C., \& Lacy, M.\ 2015, \apjl, 806, L25 
\bibitem[Stern et al.(2014)]{stern2014} Stern, D., Lansbury, G.~B., Assef, R.~J., et al.\ 2014, \apj, 794, 102
\bibitem[Symeonidis et al.(2013)]{symeonidis2013} Symeonidis, M., Vaccari, M., Berta, S., et al.\ 2013, \mnras, 431, 2317
\bibitem[Trotta (2008)]{trotta2008} Trotta, R. 2008, ConPh, 49, 71
\bibitem[Tsai et al.(2015)]{tsai2015} Tsai, C.-W., Eisenhardt, P.~R.~M., Wu, J., et al.\ 2015, \apj, 805, 90
\bibitem[Venemans et al.(2007)]{venemans2007} Venemans, B.~P., R{\"o}ttgering, H.~J.~A., Miley, G.~K., et al.\ 2007, \aap, 461, 823 
\bibitem[Wang et al.(2008)]{wang2008} Wang, R., Wagg, J., Carilli, C.~L., et al.\ 2008, \aj, 135, 1201
\bibitem[Wang et al.(2011)]{wang2011} Wang, R., Wagg, J., Carilli, C.~L., et al.\ 2011, \aj, 142, 101
\bibitem[Wright et al.(2010)]{wright2010} Wright, E.~L., Eisenhardt, P.~R.~M., Mainzer, A.~K., et al.\ 2010, \aj, 140, 1868
\bibitem[Wu et al.(2012)]{wu2012} Wu, J., Tsai, C.-W., Sayers, J., et al.\ 2012, \apj, 756, 96
\bibitem[Wu et al.(2014)]{wu2014} Wu, J., Bussmann, R.~S., Tsai, C.-W., et al.\ 2014, \apj, 793, 8
\bibitem[Wu et al.(2017)]{wu2017} Wu, J., Jun, H.~D., Assef, R.~J., et al.\ 2017, arXiv:1703.06888 
\bibitem[Wylezalek et al.(2013)]{wylezalek2013} Wylezalek, D., Galametz, A., Stern, D., et al.\ 2013, \apj, 769, 79 
\end{thebibliography}
\end{document}